# Coherently driven quantum features using a linear optics-based polarization-basis control

Byoung S. Ham
School of Electrical Engineering and Computer Science, Gwangju Institute of Science and Technology
123 Chumdangwagi-ro, Buk-gu, Gwangju 61005, South Korea
(Submitted on March 22, 2023; bham@gist.ac.kr)

**Abstract**
Quantum entanglement generation is generally known to be impossible by any classical means. According to Poisson statistics, coherent photons are not considered quantum particles due to the bunching phenomenon. Recently, a coherence approach has been applied to interpret quantum features such as the Hong-Ou-Mandel (HOM) effect, Franson-type nonlocal correlation, and delayed-choice quantum eraser, where the quantum feature is due to basis-product superposition at the cost of 50 % photon loss. For this, it has been understood that a fixed sum-phase relation between paired photons is the bedrock of quantum entanglement. Here, coherently driven quantum features of the HOM effects are presented using linear optics-based polarization-basis control. Like quantum operator-based destructive interference in the HOM theory, a perfectly coherent analysis shows the same photon bunching of the paired coherent photons on a beam splitter, whereas individual output intensities are uniform.

**Introduction**
Over the last several decades, quantum entanglement has been intensively studied for the weird quantum phenomena that cannot be obtained by classical physics [1-9]. The 'weird' quantum features are due to our limited understanding of quantum entanglement, as Einstein raised a fundamental question on nonlocal realism [1]. An intuitive answer to the impossible quantum feature by classical physics can be found in the uncontrolled tensor products of two bipartite particles, resulting in the classical lower bound in intensity correlation [10]. As shown for the self-interference of a single photon [11], the wave-particle duality has been a main issue in quantum mechanics to understand the mysterious quantum superposition [12,13]. Here, a contradictory quantum feature driven by coherence optics is presented for the 'weird' quantum features using a polarization-basis control of coherent photons. As a result, the quantum feature of photon bunching of the HOM effects is analytically demonstrated for the coincidence detection of coherent photons from a beam splitter (BS), whereas output ports show a uniform intensity. The path-length dependent coherence effect is completely removed for the coherently derived HOM effects.

Recently, a coherence approach [14-17] has been applied for entangled photon pairs generated from the spontaneous emission parametric down-conversion (SPDC) process [18,19] to interpret quantum features such as the Hong-Ou-Mandel (HOM) effects [20-22], Franson-type nonlocal correlation [23-25], and delayed-choice quantum eraser [26-29]. On the contrary to conventional particle nature-based understanding, the nonlocal quantum feature between space-like separated photons originates in phase coherence-based basis-product modification resulting from coincidence detection [15]. This phase coherence commonly applies to both distinguishable (particle nature) and indistinguishable (wave nature) characteristics of a single photon, where a specific phase relationship between the paired photons has already been derived from both HOM [14] and delayed-choice quantum eraser [16]. Such a phase relation is provided by a fixed sum phase between paired photons according to the phase-matching condition of second-order nonlinear optics [28,30]. These are the backgrounds of the present coherence approach to the coherence quantum feature using polarization-basis modification of coherent photons to understand otherwise the 'weird' quantum phenomenon.

**Results**
Figure 1(a) shows the schematic of the coherently derived quantum features using an attenuated laser via polarization-basis control. To provide random polarization bases of a single photon, the laser L is followed by a 22.5°-rotated half-wave plate (HWP). Using neutral density filters, the randomly polarized photons are maintained at a low mean photon number state, satisfying independent measurement-based statistics [31]. For the phase-matched coherent photon pairs, a set of acousto-optic modulators (AOMs) are used in both paths of the noninterfering Mach-Zehnder interferometer (NMZI), where the AOMs are synchronized and oppositely scanned each other for a given bandwidth Δ. For the polarization-basis separation of NMZI output photon pairs, an additional PBS is added to each output port of the NMZI. Due to the coincidence detection of a photon pair, two (independent) polarization-correlated photon pairs, e.g., horizontal (H)-H and vertical (V)-V photon pairs in Table 1 (color matched) are provided independently. For the proof of principle, the polarization-correlated photon pairs are tested on a BS for the quantum feature of the HOM effects.

The narrow-linewidth laser L is intensity attenuated for a low mean photon number, whose Poisson-distributed single-photon rate satisfies individual and independent statistics in measurements. For spectral bandwidth 2Δ, an AOM is inserted in each arm of the first NMZI in a double-pass scheme, as shown in the Inset of Fig. 1(a), where both AOMs are synchronized and oppositely scanned. For a given spectral bandwidth of AOMs, the diffracted photons roughly satisfy a Gaussian-like profile Δ, as shown in Fig. 1(b). To satisfy random detuning at $\pm \delta f_j$ for a $j^{th}$ photon pair, the AOM's scan rate is set to be faster than the resolving time of the single photon detector or the inverse of the mean photon number, satisfying random measurements. As a result, the output photon pairs of the NMZI result in 16 different polarization-basis combinations, whose photon characteristics are distinguishable, resulting in no interference fringe. By a followed PBS in each output port of the NMZI, transparent and reflected photons are separated into horizontal and vertical polarization groups, respectively. This linear optics-based polarization-basis separation of coherent photon pairs is critical to the present coherence method to accomplish the quantum feature, mimicking the degenerate type I entangled photon pairs from SPDC [18,19].

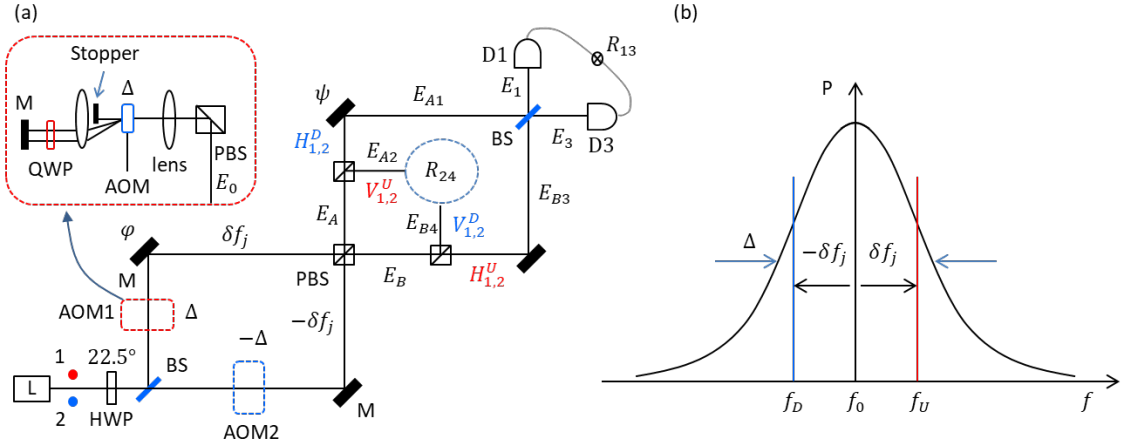

**Fig. 1. Schematic of coherence entangled photon-pair generation from an attenuated laser.** (a) Schematic of polarization-basis separation. (b) The AOM-generated spectral bandwidth of paired photons in (a). Δ is the AOM scan range. $E_0$ is the single photon's amplitude after HWP. BS: nonpolarizaing beam splitter, PBS: polarizing beam splitter. $R_{13}$: heterodyne two-photon coincidence detection.

Table 1 shows all possible polarization-basis combinations of the paired photons in Fig. 1. By definition of the coincidence detection, only doubly-bunched photons are considered with a ~1 % error rate resulting from higher-order bunched photons [31]. By the first BS of the NMZI, four possible photon-path choices are randomly allocated to each photon pair. In each photon-path choice, four different polarization-basis combinations are given randomly, resulting in a total of 16 path-polarization combinations for each pair of photons 1 and 2 (see two charts from the top). By the action of consecutive PBSs in both output paths of the first NMZI, single-path propagating photon pairs are automatically excluded from measurements (see the second and last chart). By the last PBS, both-path propagating photon pairs are separated into either orthogonally polarized or the same-polarized photon groups (see the third chart). Eventually, polarization-basis controlled photon pairs are individually tested for quantum features of the HOM effects by the last BS [20]. In Fig. 1(a), the superscript of the polarization basis indicates a corresponding up (U) or down (D) path of the first NMZI. The subscript indicates the photon number in each pair, which cannot be discernable by Poisson distribution.

**Table 1. A total of 16 possible ways to distribute photon pairs in Fig. 1(a).**

|  | Photon 1-up; Photon 2-down | | | | Photon 1-down; Photon 2-up | | | |
|---|---|---|---|---|---|---|---|---|
| Up | $H_1^U$ | $H_1^U$ | $V_1^U$ | $V_1^U$ | $H_2^U$ | $H_2^U$ | $V_2^U$ | $V_2^U$ |
| Down | $H_2^D$ | $V_2^D$ | $H_2^D$ | $V_2^D$ | $H_1^D$ | $V_1^D$ | $H_1^D$ | $V_1^D$ |

|  | Photon 1-up; Photon 2-up | | | | Photon 1-down; Photon 2-down | | | |
|---|---|---|---|---|---|---|---|---|
| Up | $H_1^U - H_2^U$ | $H_1^U - V_2^U$ | $V_1^U - V_2^U$ | $V_1^U - H_2^U$ | 0 | 0 | 0 | 0 |
| Down | 0 | 0 | 0 | 0 | $H_1^D - H_2^D$ | $H_1^D - V_2^D$ | $V_1^D - V_2^D$ | $V_1^D - H_2^D$ |

|  | Photon 1-up; Photon 2-down | | | | Photon 1-dowon; Photon 2-up | | | |
|---|---|---|---|---|---|---|---|---|
| $E_A$ | $H_2^D$ | 0 | $V_1^U - H_2^D$ | $V_1^U$ | $H_1^D$ | $H_1^D - V_2^U$ | 0 | $V_2^U$ |
| $E_B$ | $H_1^U$ | $H_1^U - V_2^D$ | 0 | $V_2^D$ | $H_2^U$ | 0 | $V_1^D - H_2^U$ | $V_1^D$ |

|  | Photon 1-up; Photon 2-up | | | | Photon 1-down; Photon 2-down | | | |
|---|---|---|---|---|---|---|---|---|
| $E_A$ | $V_2^U$ | 0 | $V_1^U - V_2^U$ | $V_1^U$ | $H_1^D$ | $H_1^D - H_2^D$ | 0 | $H_2^D$ |
| $E_B$ | $H_1^U$ | $H_1^U - H_2^U$ | 0 | $H_2^U$ | $V_2^D$ | 0 | $V_1^D - V_2^D$ | $V_1^D$ |

Table 2 shows the final sets of PBS-caused polarization-basis control for coincidence measurements in Fig. 1. By the polarization-basis separation analyzed in Table 1, the same polarization-basis sets, e.g. H-H (V-V) is independently grouped for coincidence measurements, as shown in the red- (blue-) shaded regions for detectors D1 and D3 (D2 and D4). These same-polarization-basis sets of photons satisfy the opposite frequency relation in each pair, as shown in Fig. 1(b), corresponding to the signal and idler photons from SPDC. The number '1' in the shaded regions indicates the perfect correlation between paired photons regardless of the frequency detuning in each set (see Analysis). Due to coherence, however, the cross-correlation between the orthogonal polarization-basis sets of photons also exists, as denoted by superscript δ in the off-diagonal direction. In this case, the same frequency photons are grouped in each pair. Between shaded and unshaded groups, simultaneous measurements are not allowed due to coincidence detection. The same detuned pair between D1 and D3 is also possible if two photons propagate along the same path until the last BS (see the green pairs in Table 1). This event is however eliminated by the heterodyne detection of the coincidence measurements. Thus, the present method of coherently driven quantum features using a linear optics-based polarization-basis control applies only for both shaded and unshaded regions separately. In the Analysis, the same polarization-basis groups of paired photons are considered.

**Table 2. An entangled pair chart for Fig. 1.** The subscript 'D' and 'U' indicates $-\delta f$ and $\delta f$, respectively, as shown in Fig. 1(b). '1' indicates entanglement between symmetrically (oppositely) photon detuned pairs in Fig. 1(b), whereas '$1^\delta$' is for the same frequency photons.

| Detector | | D1 | | D2 | |
|---|---|---|---|---|---|
|  | Photon | $H_1^D$ | $H_2^D$ | $V_1^U$ | $V_2^U$ |
| D3 | $H_1^U$ |  | 1 |  | $1^\delta$ |
|  | $H_2^U$ | 1 |  | $1^\delta$ |  |
| D4 | $V_1^D$ |  | $1^\delta$ |  | 1 |
|  | $V_2^D$ | $1^\delta$ |  | 1 |  |

**Analysis**

For Fig. 1, we derive coherence solutions of two-photon quantum features via coincidence detection between two output photons measured by single photon detectors D1 and D3. By definition of doubly-bunched photons and coincidence detection, simultaneous measurements between different color sets in Table 2 are not possible. At a low mean photon number, the ratio of doubly-bunched photons to single photons is ~1 % [31]. Similarly, the ratio of higher-order bunched photons to the doubly-bunched photons is ~1 % [31]. The coincidence detection eliminates both single photon and vacuum states from measurements [31]. Thus, the statistical error of coincidence measurements in Fig. 1 is ~1 %, which is negligible. This kind of statistical error is inevitable for any type of spontaneous emission process including SPDC.

From Table 2, the photon numbers 1 and 2 cannot be discernable due to identical particles given by Boson characteristics of Poisson distribution. Thus, the NMZI output photons can be represented for the $j^{th}$ pair as:

$$E_A^j = \frac{E_0}{\sqrt{2}}\left(-V^U e^{i(\varphi \pm \delta f_j t)} + H^D e^{\mp i\delta f_j t}\right), \quad (1)$$

$$E_B^j = \frac{iE_0}{\sqrt{2}}\left(H^U e^{i(\varphi \pm \delta f_j t)} + V^D e^{\mp i\delta f_j t}\right), \quad (2)$$

where $H^U$ ($H^D$) stands for the horizontal polarization basis of a UP (DOWN)-path propagating photon. Likewise, $V^U$ ($V^D$) stands for the vertical polarization basis of a UP (DOWN)-path propagating photon in the NMZI. In addition to the synchronized opposite-frequency scanning by a set of AOMs, a phase φ controller, e.g., a piezo-electric transducer (PZT) is added to the UP-path propagating photons for the first NMZI. Here, the PZT-induced phase should be dependent upon $\delta f_j$, resulting in $\varphi_j$. For simplicity, thus, PZT-induced phase is replaced by $\varphi \pm \delta f_j t \rightarrow \pm \delta f_j \tau_1(\varphi)$, where $\tau_1$ is the φ-induced time delay in the first NMZI. Due to no interaction between orthogonal polarization bases in Eqs. (1) and (2) [32,33], the corresponding mean intensities become $\langle I_A \rangle = \langle I_B \rangle = \langle I_0 \rangle$, where $I_0 = E_0 E_0^*$, and $E_0$ is the single photon amplitude.

In the second NMZI, the phase ψ is applied to $E_{A1}$ and $E_{B4}$, where these photons are from the DOWN path of the first NMZI. Like $\delta f_j \tau_1(\varphi)$, the ψ-induced phase is represented by $\delta f_j \tau_2(\psi)$, where $\tau_2$ is the ψ-induced time delay in the second NMZI. Thus, photon amplitudes used for the coincidence detection are finally represented by $E_{A1}^j = \frac{E_0}{\sqrt{2}} H^D e^{\mp i\delta f_j \tau_2}$, $E_{A2}^j = \frac{-iE_0}{\sqrt{2}} V^U e^{\pm i\delta f_j \tau_1}$, $E_{B3}^j = \frac{iE_0}{\sqrt{2}} H^U e^{\pm i\delta f_j \tau_1}$, and $E_{B4}^j = \frac{-E_0}{\sqrt{2}} V^D e^{\mp i\delta f_j \tau_2}$.

To verify the quantum feature of the two-photon correlation in Fig. 1, a conventional method of the Hong-Ou-Mandel effect is adapted for the interacting photon pairs on the BS. The amplitudes of the output photons from the BS are as follows:

$$E_1^j = \frac{1}{\sqrt{2}}\left(iE_{A1}^j + E_{B3}^j\right) = \frac{iE_0}{2}\left(H^D e^{\mp i\delta f_j \tau_2} + H^U e^{\pm i\delta f_j \tau_1}\right), \quad (3)$$

$$E_2^j = \frac{1}{\sqrt{2}}\left(iE_{A2}^j + e^{i\psi}E_{B4}^j\right) = \frac{E_0}{2}\left(V^U e^{\pm i\delta f_j \tau_1} - V^D e^{\mp i\delta f_j \tau_2}\right), \quad (4)$$

$$E_3^j = \frac{1}{\sqrt{2}}\left(E_{A1}^j e^{i\psi} + iE_{B3}^j\right) = \frac{E_0}{2}\left(H^D e^{\mp i\delta f_j \tau_2} - H^U e^{\pm i\delta f_j \tau_1}\right), \quad (5)$$

$$E_4^j = \frac{1}{\sqrt{2}}\left(E_{A2}^j + ie^{i\psi}E_{B4}^j\right) = \frac{-iE_0}{2}\left(V^U e^{\pm i\delta f_j \tau_1} + V^D e^{\mp i\delta f_j \tau_2}\right). \quad (6)$$

Thus, the corresponding mean intensities are calculated as:

$$\langle I_1 \rangle = \frac{\langle I_0 \rangle}{4}\langle \sum_j \left(H^D e^{\mp i\delta f_j \tau_2} + H^U e^{\pm i\delta f_j \tau_1}\right)\left(H^D e^{\pm i\delta f_j \tau_2} + H^U e^{\mp i\delta f_j \tau_1}\right)\rangle$$

$$= \langle \frac{I_0}{2}\rangle \langle \sum_j [1 + \cos(2\delta f_j(\tau_1 + \tau_2))]\rangle, \quad (7)$$

$$\langle I_2 \rangle = \frac{\langle I_0 \rangle}{4} \langle \sum_j (V^U e^{\pm i\delta f_j \tau_1} - V^D e^{\mp i\delta f_j \tau_2})(V^U e^{\mp i\delta f_j \tau_1} - V^D e^{\pm i\delta f_j \tau_2}) \rangle$$

$$= \langle \frac{I_0}{2} \rangle \langle \sum_j [1 - \cos(2\delta f_j(\tau_1 + \tau_2))] \rangle, \tag{8}$$

$$\langle I_3 \rangle = \frac{\langle I_0 \rangle}{4} \langle \sum_j (H^D e^{\mp i\delta f_j \tau_2} - H^U e^{\pm i\delta f_j \tau_1})(H^D e^{\pm i\delta f_j \tau_2} - H^U e^{\mp i\delta f_j \tau_1}) \rangle$$

$$= \langle \frac{I_0}{2} \rangle \langle \sum_j [1 - \cos(2\delta f_j(\tau_1 + \tau_2))] \rangle. \tag{9}$$

$$\langle I_4 \rangle = \frac{\langle I_0 \rangle}{4} \langle \sum_j (V^U e^{\pm i\delta f_j \tau_1} + V^D e^{\mp i\delta f_j \tau_2})(V^U e^{\mp i\delta f_j \tau_1} + V^D e^{\pm i\delta f_j \tau_2}) \rangle$$

$$= \langle \frac{I_0}{2} \rangle \langle \sum_j [1 + \cos(2\delta f_j(\tau_1 + \tau_2))] \rangle. \tag{10}$$

Unlike a conventional laser interference case, Eqs. (7)-(10) shows a propagation-distance proportional phase shift due simply to the opposite detuning $\pm \delta f_j \tau_k$, where $\tau_k$ is a path-length dependent transit time. Here, it should be noted that the coincidence time between the paired photons is for $\tau_1 = \tau_2$, where $2\delta f_j(\tau_1 + \tau_2) \gg 1$. Thus, $\langle 1 + \cos(2\delta f_j(\tau_1 + \tau_2)) \rangle = 1$, satisfying the uniform local intensities $\langle I_k \rangle = \frac{\langle I_0 \rangle}{2}$.

The coincidence detection between two output photons $E_1$ and $E_3$ is not like the local intensity product between Eqs. (7) and (9) because of the incompatible basis products for the same path of NMZI, as shown in Table 2:

$$\langle R_{13}(0) \rangle = \langle \sum_j E_1^j E_3^j (cc) \rangle$$

$$= \frac{\langle I_0^2 \rangle}{16} \langle \sum_j (H^D e^{\mp i\delta f_j \tau_2} + H^U e^{\pm i\delta f_j \tau_1})(H^D e^{\mp i\delta f_j \tau_2} - H^U e^{\pm i\delta f_j \tau_1})(cc) \rangle$$

$$= \frac{\langle I_0^2 \rangle}{16} H^D H^U \langle \sum_j (-e^{\mp i\delta f_j \tau_{21}} + e^{\mp i\delta f_j \tau_{21}})(cc) \rangle$$

$$= 0, \tag{11}$$

where cc is a complex conjugate, $\tau_{21} = \tau_2 - \tau_1$, and $H^k H^k = 0$. Likewise, the coincidence detection between photons $E_2$ and $E_4$ is as follows:

$$\langle R_{24}(\tau_{21}) \rangle = \langle \sum_j E_2^j E_4^j (cc) \rangle$$

$$= \frac{\langle I_0^2 \rangle}{16} \langle \sum_j (V^U e^{\pm i\delta f_j \tau_1} - V^D e^{\mp i\delta f_j \tau_2})(V^U e^{\pm i\delta f_j \tau_1} + V^D e^{\mp i\delta f_j \tau_2})(cc) \rangle,$$

$$= \frac{\langle I_0^2 \rangle}{16} V^D V^U \langle \sum_j (-e^{\mp i\delta f_j \tau_{21}} + e^{\mp i\delta f_j \tau_{21}})(cc) \rangle$$

$$= 0. \tag{12}$$

Unlike uniform local intensities in Eqs. (7)-(10), the two-photon correlation in Eqs. (11) and (12) for the coherently manipulated polarization basis shows the quantum feature of anticorrelation. In the coincidence counting module, the coincidence detection cross-correlation between the single-photon detector-generated electrical pulses whose pulse duration is a few ns. Due to the Gaussian-like spectral distribution in Fig. 1(b), the single photon-induced electrical pulse should show a similar probability distribution, resulting in a Gaussian-like cross-correlation as a function of $\tau_{21}$ [34]. The sideband oscillation of the HOM dip is from this kind cross-correlation.

**Discussion**

In Eqs. (11) and (12), the time delay $\tau_{21}$ induced by ψ and φ is in the order of $\Delta^{-1}$. Unlike local intensities in Eqs. (7)-(10), each time delay of $\tau_1$ or $\tau_2$ is in the order of the laser's coherence time which is much longer than $\Delta^{-1}$. Compared with recent coherence study of the HOM effects for entangled photons [14], Eqs. (11) and (12) show that the origin of the anticorrelaton is in the definite phase shift $\frac{\pi}{2}$ between the paired photons regardless of their spectral detuning. The random phase between photon pairs given by either Poisson statistics or the SPDC process does not deteriorate the HOM effects due to independent measurements. The same fixed sum-phase relation of the paired photons is accomplished by the first BS of the NMZI in Fig. 1. Unlike local intensities in Eqs. (7)-(10), no ensemble decoherence effect is shown in Eqs. (11) and (12) due to the selective polarization basis-products.

The linear optics-based basis selection process is the key to the quantum feature derived in Eqs. (11) and (12), resulting in the second-order quantum superposition between selected basis products of interacting photons [15]. Without coincidence detection, such a measurement-event selection process cannot be possible due to the long coherence of each photon, allowing the cross-correlation between shaded and unshaded regions in Table 2. Thus, the resolving time of a photodetector plays an important role for the coincidence detection, where this time scale must be shorter than the single photon rate. As a result, the quantum feature derived in Eqs. (10) and (11) must be limited to a microscopic regime of single photons as usually understood in quantum information science. For this, keeping a low mean-photon number is a technical requirement.

**Conclusion**

Coherently driven quantum features of the HOM effects were analyzed for the fundamental physics of quantum mechanics using linear optics-based polarization basis control of coherent photons. Unlike common understanding, the impossible quantum entanglement creation using coherent photons was analyzed for coherence manipulations of polarization-basis separation. Due to the intrinsic coherence property of mixed states, the action of the polarization-basis control by a set of PBSs resulted in an inevitable 50 % loss of measurement events. As a result, coherently induced HOM-type anticorrelation, i.e., the photon bunching phenomenon on a BS, was derived from polarization-basis modified coherent photon pairs via coincidence detection, regardless of the bandwidth. Due to the linear optics-based coherence approach, the proposed method of coherently driven HOM effects should set a new course in quantum mechanics. This work may give a step toward macroscopic entanglement generation in the future, even though such a phenomenon seems to be impossible due to mutual coherence among interacting photons at the present scope.

**References**
1. Einstein, A., Podolsky, B. &Rosen, N. Can quantum-mechanical description of physical reality be considered complete? *Phys. Rev*. **47**, 777-780 (1935).
2. Bell, J. On the Einstein Podolsky Rosen Paradox. *Physics* **1**, 195-290 (1964).
3. Clauser, J. F., Horne, M. A., Shimony, A. & Holt, R. A. Proposed experiment to test local hidden-variable theories. *Phys. Rev. Lett*. **23**, 880–884 (1969).
4. Hensen, B. *et al*. Loophole-free Bell inequality violation using electron spins separated by 1.3 kilometres. *Nature* **526**, 682–686 (2015).
5. The BIG Bell test collaboration, Challenging local realism with human choices. *Nature* **557**, 212-216 (2018).
6. Kim, T., Fiorentino M. & Wong, F. N. C. Phase-stable source of polarization-entangled photons using a polarization Sagnac interferometer. *Phys. Rev*. A **73**, 012316 (2006).
7. Jacques, V., Wu, E., Grosshans, F., Treussart, F., Grangier, P., Aspect, A., & Roch, J.-F. Experimental realization of Wheeler's delayed-choice Gedanken Experiment. *Science* **315**, 966-978 (2007).
8. Ma, X.-S., Kofler, J. & Zeilinger, A. Delayed-choice gedanken experiments and their realizations. *Rev. Mod. Phys*. **88**, 015005 (2016).
9. Horodecki R., Horodecki P., Horodecki M., & Horodecki K. Horodecki Quantum entanglement. *Rev. Mod. Phys*. **81,** 865–942 (2009).
10. Mandel, L. Photon interference and correlation effects produced by independent quantum sources. *Phys. Rev. A* **28**, 929-943 (1983).

11. Grangier, P., Roger, G. & Aspect, A. Experimental evidence for a photon anticorrelation effect on a beam splitter: A new light on single-photon interferences. *Europhys. Lett*. **1**, 173-179 (1986).
12. Dirac, P. A. M. The principles of Quantum mechanics. 4th ed. (Oxford university press, London), Ch. 1, p. 9 (1958).
13. Feynman R. P., Leighton R. & Sands M. The Feynman Lectures on Physics, Vol. III (Addison Wesley, Reading, MA (1965)
14. Ham, B. S. Coherence interpretation of the Hong-Ou-Mandel effect. arXiv:2203.13983v2 (2022).
15. Ham, B. S. The origin of Franson-type nonlocal correlation. arXiv:2112.10148v4 (2023).
16. Ham, B. S. A Coherence interpretation of nonlocal quantum correlation in a delayed-choice quantum eraser. arXiv:2206.05358v3 (2023).
17. Ham, B. S. A coherence interpretation of nonlocal realism in the delayed-choice quantum eraser. arXiv:2302.13474v2 (2023).
18. Cruz-Ramirez, H., Ramirez-Alarcon, R., Corona, M., Garay-Palmett, K. & U'Ren, A. B. Spontaneous parametric processes in modern optics. *Opt. Photon. News* **22**, 36-41 (2011), and reference therein
19. Zhang, C., Huang, Y.-F., Liu, B.-H., Li, C.-F., & Guo, G.-C. Spontaneous parametric down-conversion sources for multiphoton experiments. *Adv. Quantum Tech.* **4**, 2000132 (2021).
20. Hong, C. K., Ou, Z. Y. & Mandel, L. Measurement of subpicosecond time intervals between two photons by interface. *Phys. Rev. Lett*. **59**, 2044 (1987).
21. Lettow, R. *et al*. Quantum interference of tunably indistinguishable photons from remote organic molecules. *Phys. Rev. Lett*. **104**,123605 (2010).
22. Deng, Y.-H. *et al*. Quantum interference between light sources separated by 150 million kilometers. *Phys. Rev. Lett*. **123**, 080401 (2019)
23. J. D. Franson, Bell inequality for position and time. *Phys. Rev. Lett*. **62**, 2205-2208 (1989)
24. Kwiat, P. G., Steinberg, A. M. & Chiao, R. Y. High-visibility interference in a Bell-inequality experiment for energy and time. *Phys. Rev. A* **47**, R2472–R2475 (1993).
25. Carvacho, G. *et al*. Postselection-loophole-free Bell test over an installed optical fiber network. *Phys. Rev. Lett*. **115**, 030503 (2015).
26. Scully, M. O. & Drühl, K. Quantum eraser: A proposed photon correlation experiment concerning observation and "delayed choice" in quantum mechanics. *Phys. Rev. A* **25**, 2208-2213 (1982).
27. Kim, Y.-H., Yu, R., Kulik, S. P. & Shih, Y. Delayed "Choice" Quantum Eraser. *Phys. Rev. Lett*. **84**, 1-4 (2000).
28. Herzog, T. J., Kwiat, P. G., Weinfurter, H. & Zeilinger. A. Complementarity and the quantum eraser. *Phys. Rev. Lett*. **75**, 3034-3037 (1995).
29. DuÈrr, S., Nonn, T. & Rempe, G. Origin of quantum-mechanical complementarity probed by a `which-way' experiment in an atom interferometer. *Nature* **395**, 33-37 (1998).
30. R. W. Boyd, Nonlinear Optics, Third Edition. New York (Academic Press, 2008) pp. 79–88.
31. Kim, S. & Ham, B. S. Revisiting self-interference in Young's double-slit experiments. *Sci. Rep.* **13**, 977 (2023).
32. Hardy, L. Source of photons with correlated polarizations and correlated directions. *Phys. Lett. A* **161**, 326-328 (1992).
33. Henry, M. Fresnel-Arago laws for interference in polarized light: A demonstration experiment. *Am. J. Phys*. **49**, 690-691 (1981).
34. Nguyen, H., Duong, H. & Pham, H. Positioning the adjacent buried objects using UWB technology combine with Levenberg-Marquardt algorithm. *Info. Comm. Tech & Services* **20**, 24-32 (2022).

**Funding:** This research was supported by the MSIT (Ministry of Science and ICT), Korea, under the ITRC (Information Technology Research Center) support program (IITP 2023-2021-0-01810) supervised by the IITP (Institute for Information & Communications Technology Planning & Evaluation). BSH also acknowledges that this work was also supported by GIST GRI-2023.

**Competing Interests:** The author declares no competing interest.